\documentclass[12pt]{article}
\usepackage[english]{babel}
\usepackage[utf8x]{inputenc}
\usepackage[T1]{fontenc}

\usepackage[margin=1in]{geometry}

\usepackage{amssymb,amsmath,amsthm}
\usepackage{tikz}
\usepackage{graphicx,ctable,booktabs}
\usepackage[colorinlistoftodos]{todonotes}
\usepackage{bbm}
\usepackage{enumerate}
\usepackage{mathrsfs}
\usepackage{pdflscape}
\usepackage{tabularx}
\usepackage{subfigure}
\usepackage{natbib}
\usepackage{placeins}
\usepackage[hang,flushmargin]{footmisc}
\usepackage[capposition=top]{floatrow}
\usepackage{setspace}
\usepackage{color}
\usepackage{lscape}
\usepackage{comment}
\usepackage{dsfont}
\usepackage{verbatim}
\usepackage{multirow}
\usepackage{longtable}
\usepackage{booktabs}
\usepackage{natbib}

\usepackage{amsmath}
\usepackage{amsfonts}
\usepackage{amssymb}
\usepackage{tikz}
\usetikzlibrary{decorations.pathreplacing}
\usepackage{physics}
\usepackage{graphicx}%
\usepackage{varioref}
\usepackage[hidelinks]{hyperref}
\usepackage{enumitem}
\setlist[enumerate]{wide=0pt, widest=99,leftmargin=\parindent, labelsep=* } %\usepackage{ntheorem}
\usepackage{indentfirst}
\usepackage{threeparttable}
\usepackage{graphics}
\usepackage{verbatim} 
\usepackage{threeparttable}
\usepackage{rotating}
\usepackage[capposition=top]{floatrow}
\usepackage{caption}
\usepackage{mathtools}
\usepackage{bbm}
\usepackage[ruled,vlined]{algorithm2e}

\usepackage{xcolor,soul} % text color
\usepackage{diagbox} % added on 6/24/19
\usepackage{rotating}
\usepackage{adjustbox}
\usepackage{url}

\hypersetup{
   colorlinks=true, %set true if you want colored links
   linktoc=all,        %set to all if you want both sections and subsections linked
   linkcolor=black,  %choose some color if you want links to stand out
   urlcolor  = blue,
   citecolor = black,
}

\begin{document}
\begin{titlepage}
\onehalfspacing
\title{%
\Large Simulation of Public Cash Transfer Programs on US Entrepreneurs' Financing Constraint
%\bigskip
}

\author{{Liukun Wu} 
}
\date{September 2024}
\maketitle
\thispagestyle{empty}
\bigskip

\centerline{\bf Abstract}

\begin{singlespace}
In this paper, I conduct a policy exercise about how much the introduction of a cash transfer program as large as a Norwegian-sized lottery sector to the United States would affect startups. The key results are that public cash transfer programs (like lottery) do not increase much the number of new startups, but increase the size of startups, and only modestly increase aggregate productivity and output. The most important factor for entrepreneurs to start new businesses is their ability.

\end{singlespace}

%\let\oldthefootnote\thefootnote
%\renewcommand{\thefootnote}{\fnsymbol{footnote}}
%\footnotetext[1]{xxx }
%\let\thefootnote\oldthefootnote
\end{titlepage}

\section{Introduction}
% business policy
% purpose: get insights from theory and conjecture counterfactual for US economy
% assumption
% key contribution
% results and implications
Entrepreneurs play an important role in job creation in the US economy. Business startups contribute to about 20\% of US gross job creation (\cite{decker_role_2014}). The Small Business Administration (SBA) therefore set in place several policies dedicated to support young businesses. Some of the policies include helping small businesses win federal contracts\footnote{https://www.sba.gov/federal-contracting/contracting-assistance-programs}, providing SBA-backed loans\footnote{https://www.sba.gov/funding-programs/loans}, and various tax credit programs\footnote{https://home.treasury.gov/policy-issues/coronavirus/assistance-for-small-businesses/small-business-tax-credit-programs}. During the recent pandemic, small businesses received paycheck protection and other COVID 19 relief packages. Motivated by the recent rise in studies on nationwide lottery programs and entrepreneurship (\cite{fagereng_mpc_2021}, \cite{bermejo_entrepreneurship_2018}, \cite{kuhnetal2011aer}), I explore the question, "What will happen to entrepreneurship if the US were to run a lottery sector like Norway's?" I layout a standard Bewley model with an entrepreneur sector and a lottery sector, to understand how winning a lottery prize affects agent's consumption, saving, and occupation choices. Results of the model help to inform how effective public cash transfer programs like lottery would be to incentivize entrepreneurship, and provide lessons about its implementation. 

My model is based on \cite{kitao2008red}, in which parameters are set to match selected moments of the US economy. On top of that, there is a lottery sector, which is calibrated based on the lottery sector revenue in Norway and prize distributions in \cite{fagerengetal2016}. Agents in the economy participate in lottery every period and buy the same amount of lottery tickets. 

The theoretical insight from the model is that winning lottery prizes will increase entrepreneurs' investment for sure. On the other hand, its effect on the probability of entering business depends on abilities of would-be entrepreneurs. This is in line with empirical findings in \cite{holtzeakinetal1994a} and other recent lottery studies. 

One limitation of the model is that entrepreneurial ability is modeled using a discrete grid of four levels. There are not enough moments on entrepreneurial dynamics available to allow for a denser grid. The consequence is that there might be fewer entrepreneurs on the margin. As a result, the model might underestimate the effect of lottery prizes on the probability of entering entrepreneurship, and the total amount of entrepreneurs. 

The key contributions of this chapter is that it provides a benchmark model that can be used to study the potential effect of a lottery program in the US. Moreover, it suggests that while lottery programs are getting popular in several countries, their effectiveness is not guaranteed. The ability of marginal entrepreneurs, and the size of the lottery program, are both important in determining success. 

\section{Literature Review}
% financing constraints
This paper is related to the following three strands of literature. The first is the empirical literature that uses wealth shocks as natural experiments to study to what extent financing constraints affect the probability of starting businesses. My model sheds theoretical insight on the possible explanation for the mixed empirical results in the current literature. \cite{hurstlusardi2004jpe} challenges the prevailing opinion (\cite{evansleighton1989aer}, \cite{holtzeakinetal1994a}, \cite{holtzeakinetal1994b}) that entrepreneurs are financially constrained by showing that there is no significant relationship between household wealth and the probability to start a business over most of the wealth distribution. They further show that alternative proxies for liquidity, such as inheritance and regional house price gains, do not change the result. However, subsequent studies using inheritance and regional house price gains as wealth shocks suggest otherwise. \cite{andersennielsen2012rfs} finds receiving windfall inheritance due to sudden death of parents increases the probability of becoming an entrepreneur. \cite{corradinpopov2015} exploits regional house price variations and shows the probability of starting a new business is strongly, positively correlated with the value of home equity. And \cite{adelinoetal2015jfe} finds larger employment growth among the self-employed and small businesses in areas with steeper house price appreciation. My model shows that pure transitory income shocks do not necessarily increase the likelihood to enter entrepreneurship. It depends on the ability of entrepreneurs at the margin. Thus, it is likely that mixed results are obtained, as the distributions of entrepreneurial ability at the margin are different.

% lottery - transitory increases in disposable income 
My paper is also inspired by the recent literature on nationwide lottery programs. Compared to inheritance and home equity, lottery prizes are a better proxy for transitory increases in disposable income, especially when almost every one in the economy plays lottery games and purchases similar amount of tickets. My paper could be a benchmark theory model for settings in \cite{bermejo_entrepreneurship_2018} and \cite{cespedes_more_2021}. In addition to confirming their empirical results, my paper shows that if the US were to start a nationwide lottery program like Norway, the number of entrepreneurs in steady state will likely to be little affected. Among others, \cite{kuhnetal2011aer} uses lottery programs to study the effect of prizes on the consumption of durables and find social effects of lottery winnings.

% occupation choice
Lastly, my paper is related to the broad question of what factors determine the decision to become an entrepreneurs. Similar to \cite{evansjovanovic1989jpe} and \cite{quadrini_entrepreneurship_2000}, personal wealth affects occupation choice in my model. On the other hand, other factors, such as non-pecuniary benefits (\cite{hurstpugsley2011}) and downside insurance (\cite{hombert_can_2020}) are also shown to affect the motivation to become entrepreneurs. My paper does not touch on these dimensions, as lottery prizes bring mostly pecuniary benefits.

\section{The Model}
Based on \cite{kitao2008red}, I numerically solve an incomplete market model with an entrepreneur sector and a lottery sector, simulate lottery outcomes, and study how entry to entrepreneurship changes in response to winning a lottery. The model environment and parameters are close to \cite{kitao2008red} and match moments of the US economy.

\subsection{Endowment}
Agents are infinitely-lived. Each agent enters a period with an occupation chosen in the previous period. There are two types of occupations. The worker supplies labor for a wage in the market, and the entrepreneur owns a business, uses his managerial ability and labor, and hires capital and labor for production. I assume that entrepreneurs can only manage one business at a time.

Each agent is endowed with labor productivity $\eta$, which represents efficiency units per unit of work hours. Agents are also endowed with entrepreneurial ability $\theta$. $\eta$ and $\theta$ are assumed to be independent. Both follow a finite-state Markov process.

\subsection{Preference} 
Preferences are time-separable with a constant subjective discount factor $\beta$. Agents have expected discounted utility given as

\begin{equation}
	E_t\Big \{ \sum_{t=0}^{\infty} \beta^t u(c_t) \Big \}
\end{equation}

The utility function takes the standard CES form, $u(c)=c^{1-\sigma}/(1-\sigma)$.
$\sigma$ is the coefficient of relative risk aversion. 

\subsection{Technology and Production} 
The economy has two sectors of production. One is the entrepreneurial sector, the other is the corporate sector. \\
\textbf{Entrepreneurial sector}: each entrepreneur combines his managerial ability, capital, and labor and produces output according to the production function,
\begin{equation}
	y=f(k,l,\theta) = \theta (k^\alpha l^{1-\alpha})^v
\end{equation}
where $v<1$, capturing the entrepreneur's limited "span of control". \\
\textbf{Corporate sector}: the production function is also Cobb-Douglas except that it is constant returns to scale.
\begin{equation}
	Y=F(K,L) = AK^\alpha L^{1-\alpha}
\end{equation}

Capital in both sectors depreciate at a rate $\delta$. $\alpha$ is the capital share.
\subsection{Borrowing Sector}
The borrowing sector consists of competitive banks, which collect deposits from households and lend to both the entrepreneurial and corporate sectors. Workers face a no-borrowing constraint and cannot hold negative assets. Entrepreneurs can borrow, but their borrowing cost is higher than the corporate sector with a spread $\iota$, so that the total borrowing cost is $r_d = r+\iota$, where $r$ is the risk free rate, the rate at which banks pay for households for deposits and lend to the corporate sector. 

I also assume that due to unobserved managerial ability, entrepreneurs can only borrow up to the maximum leverage at $d$ of their net worth, and they are not allowed to default.

\subsection{Lottery Sector}
Each period, the lottery sector takes payment $\tau$ from households and distributes three types of prizes $\psi$: small, medium, and large, randomly to a subset of agents in the economy. I assume that all agents participate in the lottery, and each agent can only buy one lottery ticket at a time. 

\subsection{Government}
The government raises tax revenues to finance its expenditure $G$. A balanced budget is imposed every period. The government takes consumption tax at constant rate $\tau_c$, and an individual income tax as a function $T(I)$ of income $I$ of the agent. To approximate the US income tax system, the parametric form of the tax schedule is 
\begin{equation}
	T(I) = a_0\{I - (I^{-a_1} + a_2)^{-1/a_1}\} + \tau_I I
\end{equation}

\subsection{Household Problem}
\textbf{Timing of events}: The occupation of each agent is pre-determined from the previous period. At the beginning of each period, a pair of idiosyncratic shocks (labor productivity $\eta$ and entrepreneurial ability $\theta$), together with the outcome of lottery, are realized. Given the shocks and lottery outcomes, workers then choose their consumption and saving, while entrepreneurs make the additional decisions on the amount of labor and capital. Production then takes place in both the corporate and entrepreneurial sectors, followed by factor payments and loan repayments. Workers and entrepreneurs pay income taxes. There is no aggregate uncertainty. Lottery prizes are independent and identically distributed across agents and over time. \\

\textbf{Optimization problem}. Given assets $a$, labor productivity $\eta$, managerial ability $\theta$, lottery prize $\psi$, and prices $r,w$, households choose consumption, savings, and next period's occupation to maximize the present value of discounted utility. Denote worker's and entrepreneur's value functions by $V^W$ and $V^E$ respectively, the recursive problem is defined as,\\

\textbf{Worker's problem}
\begin{equation}
	V^W(a,\theta,\eta,\psi) = \max_{c,a',i} \{u(c) + i\beta EV^W(a',\theta',\eta',\psi') + (1-i)\beta EV^E(a',\theta',\eta',\psi')   \}
\end{equation}
subject to
\begin{align}
	(1+\tau_c)c + a' = \eta w + (1+r)a +\psi - \tau - T(I) \\
	I = \eta w + ra + \psi - \tau \\
	a' \geq 0, \quad c\geq 0, \quad i \in \{0,1\}
\end{align}
where $i$ is an indicator function that equals 1 if the agent is a worker in the next period and 0 otherwise. The worker's taxable income consists of labor income, capital income from savings, and net lottery income.\\

\textbf{Entrepreneur's problem}
\begin{equation}
	V^E(a,\theta,\eta,\psi) = \max_{c,a',i} \{u(c) + i\beta EV^W(a',\theta',\eta',\psi') + (1-i)\beta EV^E(a',\theta',\eta',\psi')   \}
\end{equation}
subject to
\begin{align}
	(1+\tau_c)c + a' = \pi^E(a,\eta,\theta,\psi) + \psi - \tau \\
	a' \geq 0, \quad c\geq 0, \quad i \in \{0,1\}
\end{align} 
where $\pi^E$ is the after tax income to the entrepreneur and determined as,
\begin{equation}
	\pi^E(a,\theta,\eta,\psi) = \max_{k,n}\{f(k,n,\theta) 
	+ (1-\delta)k - (1+\tilde{r})(k-a) - w\max \{n-\eta,0\} - T(I)   \}
\end{equation}
where
\begin{align}
	I &= f(k,n,\theta) - \delta k - \tilde{r}(k-a) - w\max\{n-\eta,0\} - \tau + \psi \\
	k &\leq (1+d)(a-\tau+\psi)
\end{align}
and 
\begin{equation*}
	\tilde{r} = 
	\begin{cases}
		r \qquad & \text{if}\ k \leq a \\
		r + \iota \qquad & k > a
	\end{cases}
\end{equation*}
Entrepreneur's income consists of output from production, net of depreciation cost, net labor cost, and loan repayments. If entrepreneur is a net borrower, $k>a$, then he pays interest at rate $r+\iota$ for the borrowing. If he is a net saver, then he earns interest at rate $r$ on the remaining assets $a-k$.

Winning a lottery prize, $\psi$, adds to entrepreneurs' current period income $I$ and relaxes borrowing constraint.

% recalculate first order conditions

\subsection{Stationary Competitive Equilibrium}
At the beginning of each period, the state vector is $s=(a,\eta,\theta,\psi,\epsilon)$, i.e. asset holdings, labor productivity, entrepreneurial ability, lottery prizes, and occupation $\epsilon \in \{W,E\}$.
Let $a\in \mathbb{A},\eta \in \mathbb{H},\theta \in \Theta,\psi \in \Psi$  and $\epsilon \in \xi$. Denote the entire state space by $\mathbb{S} = \mathbb{A}\times\mathbb{H}\times \Theta\times \Psi \times \xi$. An equilibrium consists of prices $\{r,w\}$, policy functions of workers and entrepreneurs, government tax system, intermediaries, lottery sector, value functions and the distribution of agents over state space $\mathbb{S}$ given by $\Phi(s),s\in \mathbb{S}$, such that
\begin{enumerate}
	\item Given the prices and tax, the policy functions solve the household optimization problem for each state $s$.
	\item The prices satisfy the marginal productivity conditions, i.e. $r=F_K(K,L)-\delta$, and $w=F_N(K,L)$.
	\item The intermediary sector is competitive. Banks receive deposits from households and pay interest $r$, and offer loans to the corporate sector and entrepreneurs at rate $r$ and $r+\iota$ respectively.
	\item Capital and labor markets clear:
	\begin{align*}
		K + \int k d\Phi(s) &= \int a d\Phi(s) \\
		N + \int n d\Phi(s) &= \int \eta(s) d\Phi(s)
	\end{align*}
	\item The lottery sector clears total rewards with revenue:
	\begin{equation*}
		\tau = \int \psi(s) d\Phi(s)
	\end{equation*}
	\item Government budget is balanced:
	\begin{equation}
		G = \int [\tau_c c(s) + T(I(s))] d\Phi(s)
	\end{equation}
	\item The distribution $\Phi$ is time-invariant. The law of motion for the distribution of agents over state space $\mathbb{S}$ satisfies
	\[ \Phi = Q_\Phi(\Phi) \]
	where $Q_\Phi$ is the one-period transition operation on the distribution.
	
\end{enumerate}

\section{Calibration} 
% https://stats.oecd.org/Index.aspx?DataSetCode=REV tax revenue as % of GDP
% http://world.tax-rates.org/norway/sales-tax sales tax 25%
% https://www.nordisketax.net/main.asp?url=files/nor/eng/i07.asp ordinary tax 23%
% https://data.oecd.org/norway.htm
I will be calculating and comparing two steady state economies, one with and one without the lottery sector. The benchmark economy is without the lottery sector and resembles the current US economy. Table \ref{ch3_params} gives the set of a priori parameters and are based on \cite{kitao2008red}. These parameters are set to match moments such as capital-output ratio, government expenditures, and entrepreneur dynamics.

\begin{table}[h] \centering%
	\centering
	\caption{A Priori Parameters}%
	\begin{tabular}{ lll}
		\hline
		Parameter & Description & Values \\
		\hline
		Preference && \\
		$\sigma$ & relative risk aversion & 2.0 \\
		$\beta$  & discount factor & 0.9575 \\
		Production technology && \\
		$\alpha$ & capital share in the corporate sector & 0.36 \\
		$v$ & entrepreneur's span of control & 0.88 \\
		$\delta$ & depreciation rate of capital & 0.06 \\
		Labor and entrepreneurship && \\
		$\eta$ & efficiency unit & Appendix \\
		$\theta$ & entrepreneurial ability & Appendix \\
		Tax system&&\\
		$\tau_c$ & consumption tax rate & 0.0567 \\
		$\tau_I$ & proportional tax rate on income & 0.0316 \\
		$\{a_0,a_1,a_2\}$ & nonlinear tax schedule & $\{0.258,0.768,0.438\}$\\
		Intermediary sector &&\\
		$\iota$  & loan premium & 0.05\\
		$d$ & maximum leverage ratio & 0.5\\
		\hline
	\end{tabular}
	\label{ch3_params}
\end{table}

%\noindent \textbf{Tax}: in Norway, the sales tax rate (VAT) in Norway is 25\% \footnote{http://world.tax-rates.org/norway/sales-tax}. Tax rate on oridinary income in 2018 is 23\% \footnote{https://www.nordisketax.net/main.asp?url=files/nor/eng/i07.asp}.

\begin{figure}[h]
	\begin{center}
		\includegraphics[width=\textwidth]{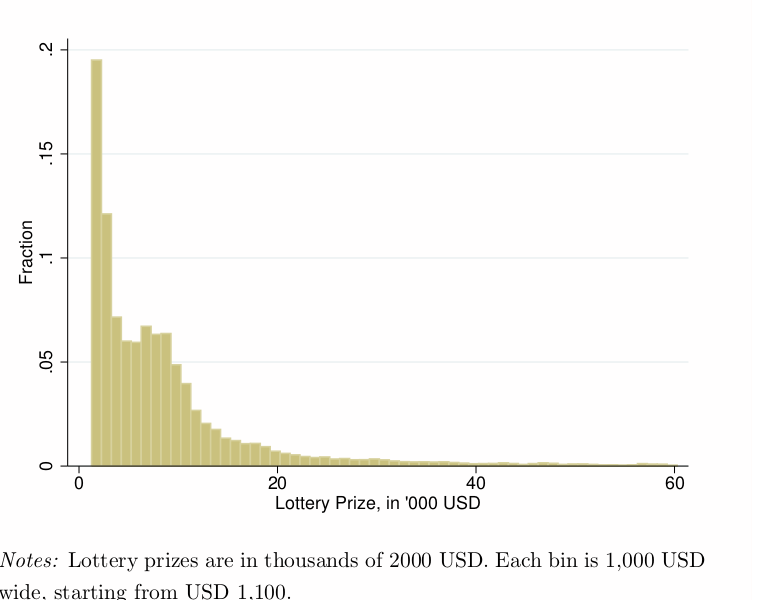}
		\caption{Lottery Prize Distribution - \cite{fagerengetal2016}}
		\label{ch3_fig1}
	\end{center}
\end{figure}

\textbf{Lottery Sector}: in order to set up the winning probabilities, I use the distribution of lottery prizes, shown in Fig \ref{ch3_fig1}, from \cite{fagerengetal2016}, as it is the only paper that provides detailed description of the lottery sector and prize distributions. To simplify the computation, I discretize the lottery prize distribution as follows. First, the sizes come in small, medium, and large, as the prize distribution features three intervals (0 to 20,000, 20,000 to 40,000, 40,000 to 60,000) with distinctly different winning probabilities. Then, I set the relative prize magnitudes to be 1, 3, and 6, as the largest prize is about 60,000 USD.

Based on \cite{fagerengetal2016}, the fraction of winners is $0.5\%$. I make 94\% of the winners win the small prize, corresponding to the total fraction of winners with prizes less than 20,000 USD in Figure \ref{ch3_fig1}. 5\% of the winners win a medium prize, corresponding to the total fraction of winners with prizes between 20,000 to 40,000 USD, and 1\% of the winners win a large prize, corresponding to the total fraction of winners with prizes between 40,000 to 60,000 USD. The size of the smallest prize is to be calibrated in order for total prize amount to match total lottery revenue.

To set the lottery ticket price, I refer to the Norwegian Gambling Authority's statistics on gambling revenue, which states that in 2017, gross revenue in the regulated gambling market is 43.7 billion NOK.\footnote{https://lottstift.no/pengespel/pengespelstatistikk/norsk-pengespelmarknad-i-2017/} This is about 1.32\% of Norwegian GDP (3,299 billion NOK) in 2017. The average amount spent on regulated gambling market is 10,600 NOK, which is abou 3.3\% of the average net disposable income in Norway. Assuming that each adult spend the same amount on gambling, I back out the population that participate in gambling to be around 4.1 million - about 80\% of the Norwegian population. This almost includes every working adult, or individuals above 18 years old. Thus, $\tau$ is set to match 1.32\% of total output in the model. Assume that the US has a lottery sector that's identical to Norway's, $\tau$ should be set to match 1.32\% of the output.

% https://lottstift.no/pengespel/pengespelstatistikk/norsk-pengespelmarknad-i-2017/
% 43.7 billion NOK 2017 gaming revenue
% 3299 billion NOK gdp

\begin{table}[h] \centering%
	\centering
	\caption{Lottery Sector Parameters}%
	\begin{tabular}{ lll}
		\hline
		Parameter & Description & Values \\
		\hline
		$\tau$ & lottery ticket price & 0.0292\footnote{In equilibrium, this is about 0.24\% of a median worker's income.} \\
		$p_1$  & non-winning probability & 0.9950 \\
		$\{p_2,p_3,p_4\}$ & winning probability & $\{0.0047,0.00025,0.00005\}$  \\
		$\psi$ & single prize unit & 5.08 \\
		$\{\psi_1,\psi_2,\psi_3,\psi_4\}$& relative prize magnitude & $\{0,1,3,6\}$\footnote{The smallest prize is about 42\% of a median worker's income, and the largest prize is about 2.5 times of a median worker's income.}\\
		\hline
	\end{tabular}
	\label{ch3_lottery}
\end{table}

\section{Benchmark Economy}
Table \ref{ch3_benchmark} compares the moments generated by the benchmark economy without the lottery sector and US data. Moments about aggregate capital to output ratio, entrepreneur investment, entry and exit to entrepreneurship, and taxation are fairly close to the US economy. We will compare this steady state with the one with a lottery sector later.

\begin{table}[h] \centering%
	\centering
	\caption{Benchmark Model Moments and Data}%
	\begin{tabular}{ lcc}
		\hline
		& US Data & Model \\
		\hline
		capital-output ratio & 2.65 & 2.60 \\
		government expenditure/GDP & 18\% & 17.58\% \\
		income tax/total tax revenue & 65\% & 64.57\% \\
		fraction of entrepreneurs & 12\% & 13.74\% \\
		share of entrepreneur's income & 27\% & 26.79\% \\
		exit rate & 20\% & 21.30\% \\
		\hline
	\end{tabular}
	\label{ch3_benchmark}
\end{table}

\begin{figure}[h]
	\begin{center}
		\includegraphics[width=\textwidth]{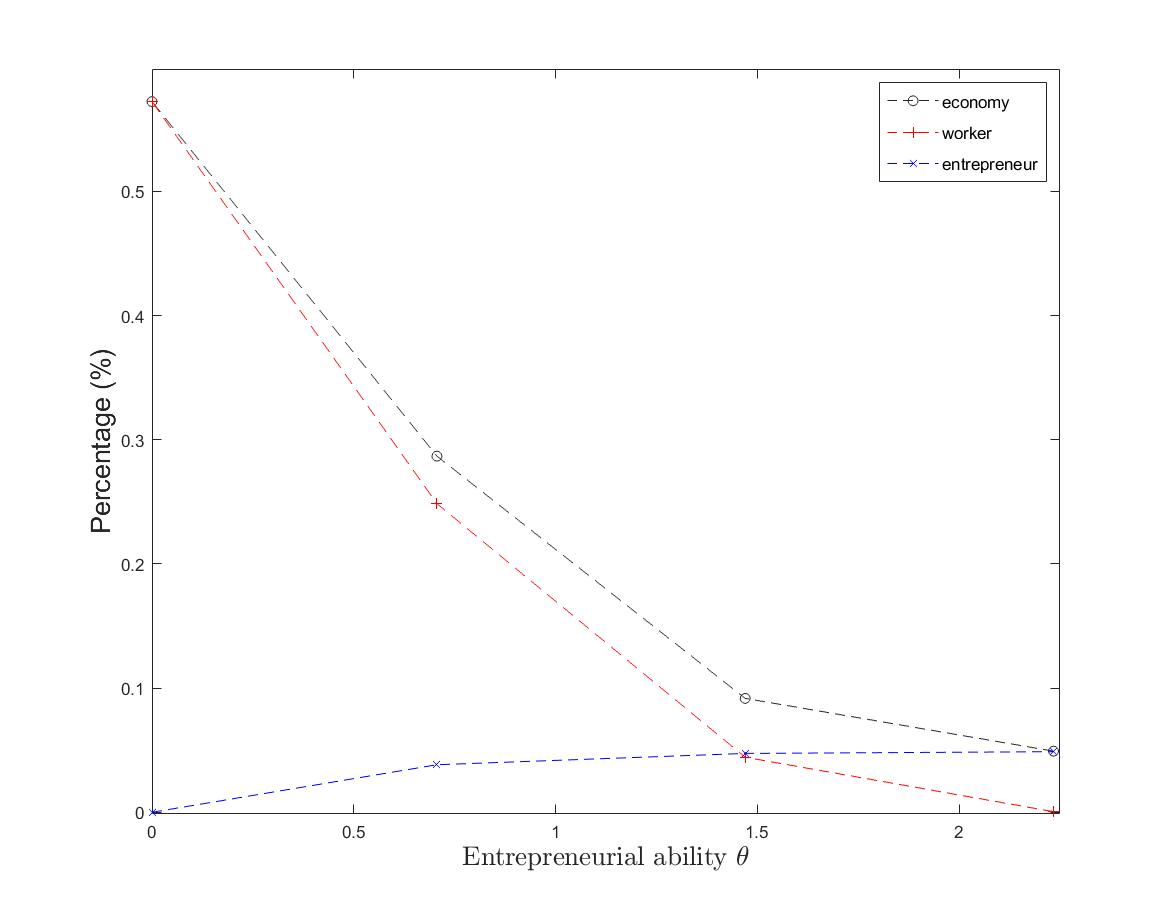}
		\centering
		\caption{Benchmark model: distribution of abilities $\theta$}
		\label{ch3_fig2}
	\end{center}
\end{figure}

% ability distribution, investment, labor, larger grid
We then examine the composition of entrepreneurs. Figure \ref{ch3_fig2} shows the distribution of entrepreneurial abilities $\theta$ in the economy, workers and entrepreneurs respectively. Agents with the highest ability are almost all entrepreneurs, and the lowest abilities are all workers. The percentage of entrepreneurs increases with ability.

Table \ref{ch3_benchmark_theta} gives a detailed description of entrepreneurs' activities by ability. Leverage ratio is defined as the percentage of assets borrowed from the intermediaries to put into investment. High ability entrepreneurs invest more into their projects and take on the maximum leverage allowed.

\begin{table}[h] 
	\centering
	\caption{Benchmark model: entrepreneurial activities by ability $\theta$}%
	\begin{tabular}{cccccc}
		\hline
		$\theta$ & \% in pop. & \% entrep. & avg. investment & avg. assets & avg. leverage ratio \\
		\hline
		0.000 & 0.00\%  & 0.00\% & & & \\
		0.706 & 3.81\%  & 27.71\% & 3.15 & 6.84 & 1.15\%  \\
		1.470 & 4.79\%  & 34.89\% & 8.78 & 8.05 & 20.39\% \\
		2.234 & 5.14\%  & 37.41\% & 16.26 & 10.84 & 50.00\% \\
		\hline
	\end{tabular}
	\label{ch3_benchmark_theta}
\end{table}

\begin{figure}[h]
	\begin{center}
		\includegraphics[width=\textwidth]{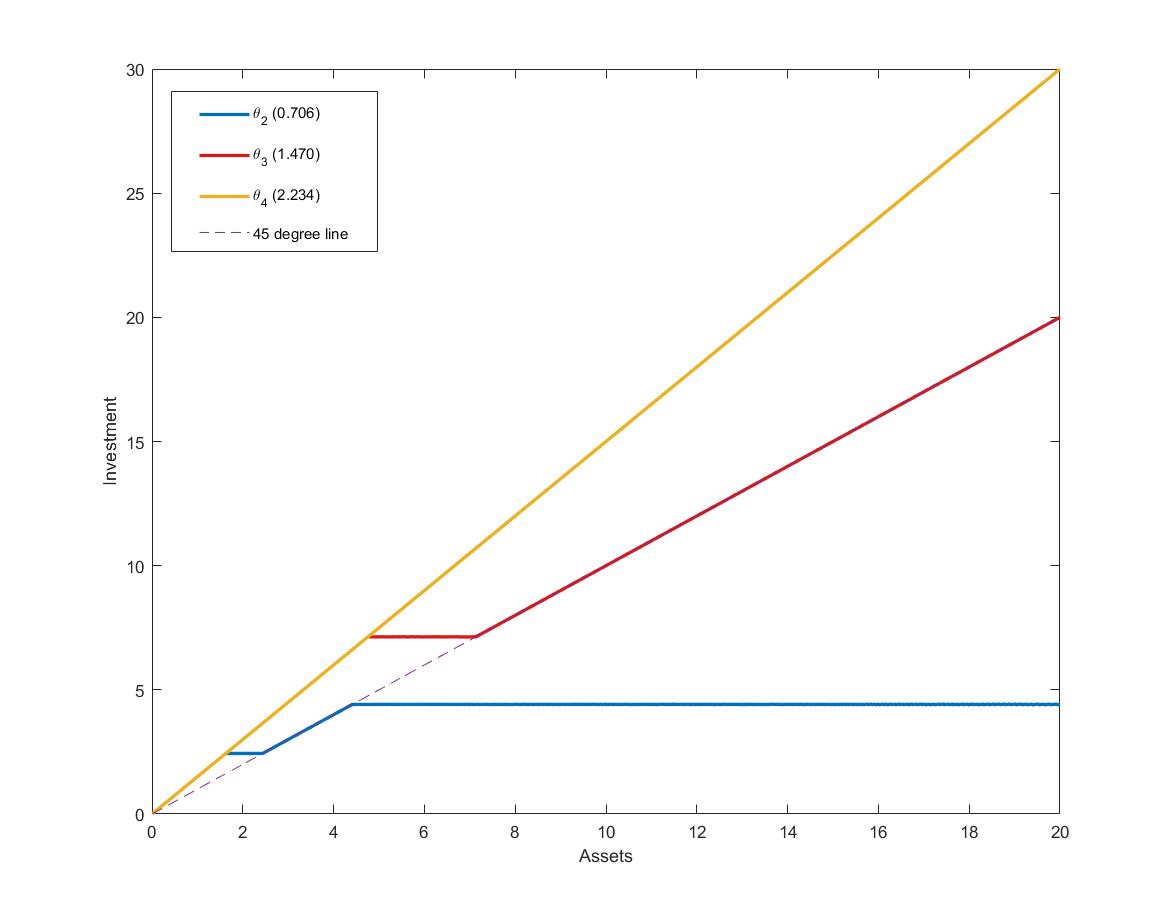}
		\centering
		\caption{Benchmark model: investment by asset and ability}
		\label{ch3_fig3}
	\end{center}
\end{figure}

Figure \ref{ch3_fig3} shows entrepreneurs' investments by asset and ability. With the decreasing scales to return technology, low ability entrepreneurs ($\theta_2$) reach the optimal firm size at a small asset level. Medium ability entrepreneurs ($\theta_3$) with smaller assets borrow up to the borrowing limit, but those with larger assets do not borrow beyond their assets due to the borrowing premium $\iota$. High ability entrepreneurs expand until they reach the borrowing constraint.

\begin{figure}[h]
	\begin{center}
		\includegraphics[width=\textwidth]{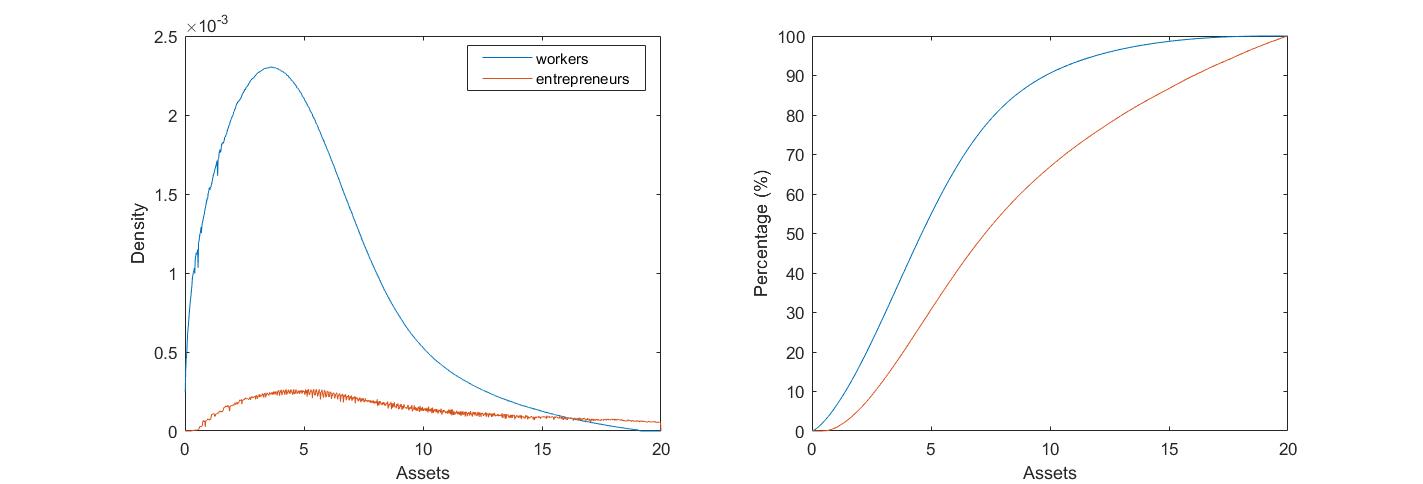}
		\centering
		\caption{Benchmark model: density and CDF of assets by occupation}
		\label{ch3_fig4}
	\end{center}
\end{figure}

% wealth, asset comparison, distributions
Entrepreneurs own a higher level of assets. Figure \ref{ch3_fig4} compares the asset distribution between workers and entrepreneurs. The left panel shows that the density of workers is concentrated in asset values less than 10 (maximum asset holdings is 20). 

\section{Lottery Experiments}
I then solve the steady state for an economy with a lottery sector, using parameters in both Tables \ref{ch3_params} and \ref{ch3_lottery}. First, I compare moments of aggregate output, investment, consumption, and entrepreneur dynamics with the benchmark economy. Then, I simulate a panel of households, and run a regression to test whether the probability to enter entrepreneurship increases after winning a lottery prize. Lastly, I test different lottery prize structures to see if the distribution of prizes affect agent's behavior in equilibrium.

\subsection{Steady State Results}
% table of eqm moments + aggregate output, capital, consumption
Table \ref{ch3_table1} compares steady state moments between the lottery and benchmark economies and calculates the percentage change. The two economies are very similar. The lottery economy has a slightly higher aggregate output and capital, and lower consumption. It also has fewer entrepreneurs, less income and investment made by entrepreneurs, and higher exit rate. The magnitudes of change are all within 5\%.

\begin{table}[h] \centering%
	\centering
	\caption{Steady State Moments}%
	\begin{tabular}{ lccc}
		\hline
		& Lottery & Benchmark & \% change\\
		\hline
		aggregate output &  2.18  & 2.17  &  +0.46\% \\
		aggregate capital & 5.79   & 5.65 &  +2.48\%  \\
		aggregate consumption & 1.37  & 1.37 & -0.64\% \\
		capital-output ratio & 2.65 & 2.60 & +1.92\% \\
		government expenditure/GDP & 17.54\% & 17.58\%  & -0.23\% \\
		income tax/total tax revenue & 64.77\% & 64.58\%  & +0.29\% \\
		fraction of entrepreneurs & 13.52\% & 13.74\% & -1.60\% \\
		share of entrepreneur's income & 26.24\% & 26.79\%  & -2.05\% \\
		share of entrepreneur's asset  & 20.71\%	& 21.32\% &-2.86\%	\\
		share of entrepreneur's investment & 23.63\%	&  24.39\%	  & -3.12\% \\
		exit rate & 21.40\% & 21.30\% & +0.47\% \\
		
		\hline
	\end{tabular}
	\label{ch3_table1}
\end{table}

% entry, investment, leverage by ability - compare with benchmark
I then look at entrepreneur's activities in more detail and compare it with the benchmark economy. Table \ref{ch3_table2} lists entrepreneur's composition, investment, asset, and leverage ratio by ability, and these results are similar to the benchmark results, within 3\% of difference. There is a slight drop in the number of entrepreneurs, but each type of entrepreneurs on average has slightly larger investment and asset.

\begin{table}[h] 
	\centering
	\caption{Lottery economy: entrepreneurial activities by ability $\theta$}%
	\begin{tabular}{cccccc}
		\hline
		$\theta$ & \% in pop. & \% entrep. & avg. investment & avg. assets & avg. leverage ratio \\
		\hline
		0.000 & 0.00\%  & 0.00\% & & & \\
		0.706 & 3.77\%  & 27.91\% & 3.20 & 6.99 & 1.21\%  \\
		1.470 & 4.73\%  & 34.98\% & 8.90 & 8.17 & 19.98\% \\
		2.234 & 5.02\%  & 37.11\% & 16.47 & 10.95 & 50.00\% \\
		\hline
	\end{tabular}
	\label{ch3_table2}
\end{table}

\begin{figure}[h]
	\begin{center}
		\includegraphics[width=\textwidth]{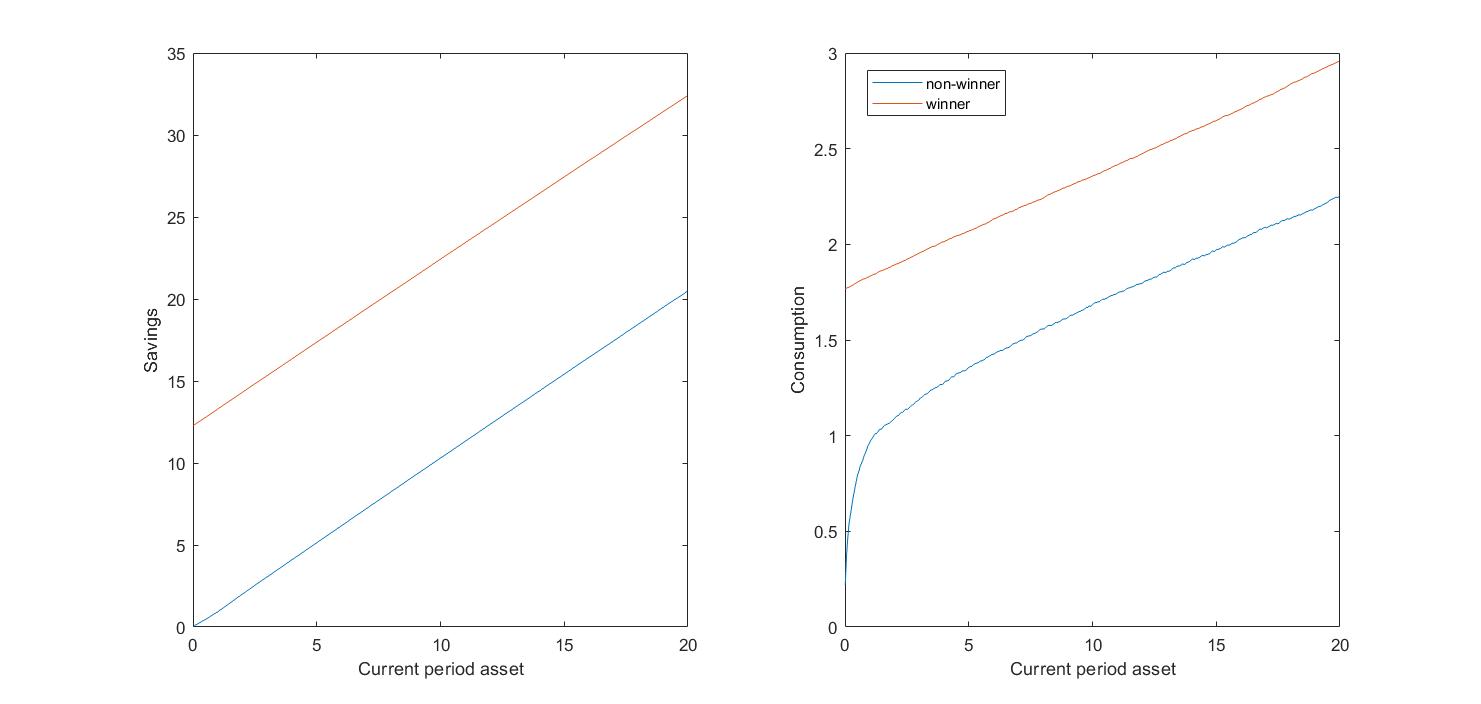}
		\centering
		\caption{Lottery economy: savings and consumption policy by winning status}
		\label{ch3_fig5}
	\end{center}
\end{figure}

% winners vs non-winners: savings, consumption, investment (describe occupation policy)
Within the lottery economy, I compare the behavior between winners and non-winners. Figure \ref{ch3_fig5} shows the consumption and savings policy of winners and non-winners. For the ease of illustration, I take the mean across prizes at each asset level for winners. In line with the permanent income hypothesis, winners save a fraction of the prizes and consumes the rest, hence having higher consumption and saving levels than non-winners. 

\begin{figure}[h]
	\begin{center}
		\includegraphics[width=\textwidth]{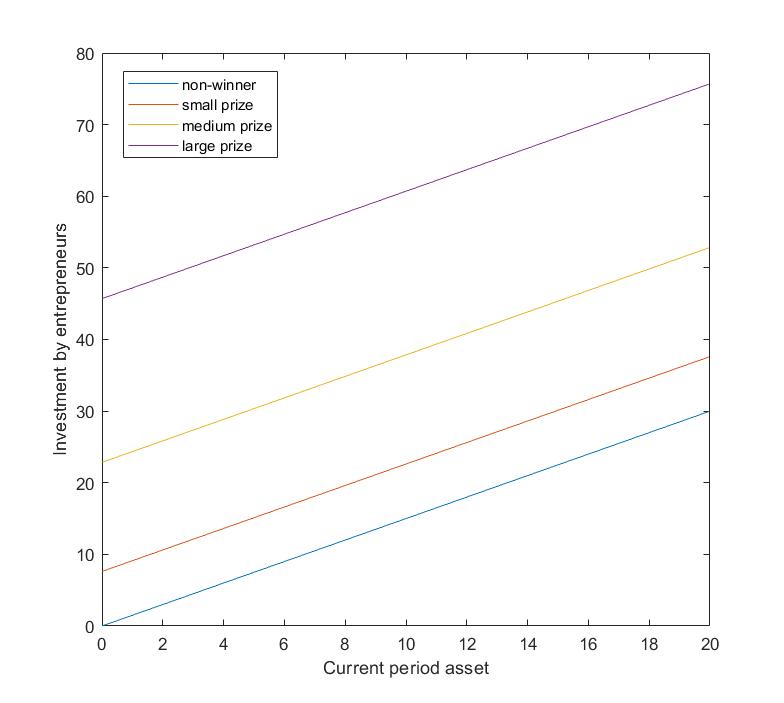}
		\centering
		\caption{Lottery economy: entrepreneur's investment policy by winning status}
		\label{ch3_fig6}
	\end{center}
\end{figure}

Figure \ref{ch3_fig6} shows entrepreneurs' investment policy across different prizes. Entrepreneurs who win prizes invest much more than those who do not, and those who win a larger prize invest more than those who win a smaller prize. Lottery prizes increases the collateral entrepreneurs can put up, relaxes the borrowing constraint, hence allowing entrepreneurs to borrow more capital. Table \ref{ch3_table3} compares the average leverage ratio across different prizes. As lottery prize gets larger, the more capital entrepreneur is allowed to borrow, and the more likely that his business will reach optimal size give his ability. Thus, leverage ratio decreases with the size of prizes.

\begin{table}[h] 
	\centering
	\caption{Lottery economy: leverage across prizes}%
	\begin{tabular}{lcccc}
		\hline
		prize & zero & small & medium & large \\
		\hline
		ave. leverage & 25.91\%  & 18.73\% & 18.55\% & 18.55\% \\
		\hline
	\end{tabular}
	\label{ch3_table3}
\end{table}

At the extensive margin, agents with abilities above certain threshold will become an entrepreneur even with the minimum amount of assets (0.05), when they win a lottery prize. On the other hand, for non-winners, they would have to accumulate enough assets to become entrepreneurs and have a high enough ability.

\subsection{Simulation}
% consider regression vs steady state results
Given the consumption and savings policy functions, and occupation decisions in steady state, I simulate a panel of 400,000 households over 200 time periods\footnote{Average asset in the economy converge after 100 periods.}, and run the following regressions for the last period,
\begin{equation}
	Entrepreneur_{i,t} = \beta_1 \phi_{i,t} + \beta_2 a_{i,t-1} + \epsilon_{i,t}
\end{equation}

\begin{equation}
	k_{i,t} = \beta_1 \phi_{i,t} + \beta_2 a_{i,t-1} + \epsilon_{i,t}
\end{equation}

The first regression tests the hypothesis that winning a lottery prize will increase agent's propensity of becoming an entrepreneur. And the second regression tests whether winning a lottery prize increases entrepreneur's investment, conditional on being an entrepreneur. I do not include ability ($\theta$) in the regressions as $\theta$ is not directly observable outside the model context.

\begin{table}[h]
	\centering
	\caption{Regression: entrepreneur decision, consumption, and investment}
	\def\sym#1{\ifmmode^{#1}\else\(^{#1}\)\fi}
	\begin{tabular}{l*{3}{c}}
		\hline\hline
		&\multicolumn{1}{c}{Entrepreneur}&\multicolumn{1}{c}{Consumption}&\multicolumn{1}{c}{Investment}\\
		\hline
		
		prize, $\psi$       &     0.003         &      0.047\sym{***}&       0.418\sym{***}\\
		&   (0.010)         &   (0.004)         &    (0.095)         \\
		
		wealth, $a$ &      0.046\sym{***}&      0.077\sym{***}&       0.921\sym{***}\\
		&  (0.000)         &  (0.000)         &   (0.003)         \\
		
		%			ability, $\theta$     &                     &                     &       17.70\sym{***}\\
		%			&                     &                     &    (0.085)         \\
		\hline
		\(N\)       &      400000         &      400000         &       63440         \\
		\(R^{2}\)   &       0.068              &       0.600         &       0.810         \\
		\hline\hline
	\end{tabular}
	\begin{tablenotes}
		\item \footnotesize Standard errors in parentheses, \sym{*} \(p<0.05\), \sym{**} \(p<0.01\), \sym{***} \(p<0.001\)
	\end{tablenotes}
	\label{ch3_table4}
\end{table}

Table \ref{ch3_table4} displays the regression results. Surprisingly, winning a lottery prize does not significantly influence the decision to become an entrepreneur. On the other hand, higher asset levels increase the probability of entrepreneurship. This is because when ability is not included in the regression, wealth becomes a proxy for ability. A higher wealth level not only loosens financing constraints, but is also associated with higher ability. Moreover, as seen in the model, ability is the dominant factor for entrepreneurship. Thus, the extensive margin depends on wealth levels, but not lottery prizes.

On the other hand, winning a lottery prize significantly increases entrepreneur's investment. Getting \$1 more prize increases investment by \$0.42. For agents with high enough ability ($\theta$) to start businesses, winning a lottery prize increases their borrowing capacity and allows them to increase investment. As a robustness check, winning a lottery prize significantly increases agent's consumption as expected.

\subsection{Different Lottery Prize Structures}
% only small, only large
As robustness checks, I experiment with different lottery prize structures. I keep the lottery ticket price, $\tau$, unchanged, and vary the size of lottery prizes. Table \ref{ch3_table5} shows two examples. The small prize structure keeps the percentage of winner unchanged at 0.5\%, but every winner is getting the same prize, 5.84, which is about the size of the smallest prize in the lottery economy. The large prize structure gives every winner a large prize, 58.4, that is ten times the smallest prize in the lottery economy, but concentrates winners to 0.05\% of the population. These two examples are used to examine if the size of lottery prizes and the distribution of prizes has an effect on the decision to entrepreneurship. I simulate the same panel of households with the same sequence of shocks\footnote{I have to regenerate the shocks for lottery prizes because the number of prizes is different.} and run the same regressions.

\begin{table}[ht] \centering%
	\caption{Alternative Lottery Structure}%
	\begin{tabular}{ lll}
		\hline
		Parameter & Description & Values \\
		\hline
		Small prize && \\
		\hline 
		$\{\phi_1,\phi_2\}$ & prizes& $\{0,\phi\},\phi=5.84$  \\
		$\{p_1,p_2\}$ & prize probability & $\{0.9950,0.005\}$\\
		\hline
		Large prize && \\
		\hline 
		$\{\phi_1,\phi_2\}$ & prizes& $\{0,\phi\},\phi=58.4$  \\
		$\{p_1,p_2\}$ & prize probability & $\{0.9995,0.0005\}$\\
		\hline 
	\end{tabular}
	\label{ch3_table5}
\end{table}

Table \ref{ch3_table6} and Table \ref{ch3_table7} shows the results for the small and large prize structures respectively. I obtain the same result that winning a lottery prize only increases investment for entrepreneurs at the intensive margin, but not the decision to become an entrepreneur (extensive margin).

\begin{table}[h]
	\caption{Entrepreneur decision, consumption, and investment - Small Prize}
	\def\sym#1{\ifmmode^{#1}\else\(^{#1}\)\fi}
	\begin{tabular}{l*{3}{c}}
		\hline\hline
		&\multicolumn{1}{c}{Entrepreneur}&\multicolumn{1}{c}{Consumption}&\multicolumn{1}{c}{Investment}
		\\
		\hline
		
		prize, $\psi$       &     0.003         &      0.074\sym{***}&       0.615\sym{***}  \\
		
		&    (0.011)         &   (0.005)         &    (0.066)         \\
		
		wealth, $a$ &       0.105\sym{***}&       0.106\sym{***}&       1.000\sym{***} \\
		&  (0.001)         &  (0.000)         &   (0.003)    \\
		
		%		ability, $\theta$     &                     &                     &      10.000\sym{***} \\
		%							&                     &                     &    (0.047)      \\
		\hline
		\(N\)      &      400000         &      400000         &       62606  \\
		\(R^{2}\)   &      0.050               &       0.526         &       0.857     \\
		\hline\hline
	\end{tabular}
	\begin{tablenotes}
		\item \footnotesize Standard errors in parentheses, \sym{*} \(p<0.05\), \sym{**} \(p<0.01\), \sym{***} \(p<0.001\)
	\end{tablenotes}
	\label{ch3_table6}
\end{table}

\begin{table}[h]
	
	\caption{Entrepreneur decision, consumption, and investment - Large Prize}
	\def\sym#1{\ifmmode^{#1}\else\(^{#1}\)\fi}
	\begin{tabular}{l*{3}{c}}
		\hline\hline
		&\multicolumn{1}{c}{Entrepreneur}&\multicolumn{1}{c}{Consumption}&\multicolumn{1}{c}{Investment}
		\\
		\hline
		
		prize, $\psi$       &    0.001         &      0.059\sym{***}&       0.381\sym{**}   \\
		
		&   (0.004)         &   (0.007)         &     (0.133)           \\
		
		wealth, $a$ &      0.026\sym{***}&      0.064\sym{***}&       0.852\sym{***}\\
		&  (0.000)         &  (0.000)         &   (0.004)   \\
		
		%		ability, $\theta$     &                     &                     &      26.03\sym{***} \\
		%		&                     &                     &    (0.124)      \\
		\hline
		\(N\)      &      400000         &      400000         &       63913  \\
		\(R^{2}\)   &      0.050               &       0.664         &       0.753     \\
		\hline\hline
	\end{tabular}
	\begin{tablenotes}
		\item \footnotesize Standard errors in parentheses, \sym{*} \(p<0.05\), \sym{**} \(p<0.01\), \sym{***} \(p<0.001\)
	\end{tablenotes}
	\label{ch3_table7}
\end{table}

\section{Results Discussion}
% what has happened here
The lottery sector randomly redistributes capital. It takes a small amount of income from each agent and concentrates the combined income on a much smaller sub-population, the 0.5\% winners in the economy. High ability agents might thus receive a large enough income to either overcome the financing constraint at entry, or expand firm size to the optimum given a higher borrowing limit. I observe both effects in the lottery economy. On the margin, high ability agents ($\theta > 1.47$) choose entrepreneurs as their next period's occupation after winning a prize, even though they are at a low asset level. Moreover, agents invest \$0.4 more for every \$1 more prize given.

However, entrepreneurial ability is the dominant factor in deciding entrepreneurship. Even though marginal entrepreneurs can use lottery prizes to overcome financing constraints, their abilities decide how profitable their businesses will be. As a result, winning lottery prizes may or may not increase entry to entrepreneurship, depending on the distribution of ability for marginal entrepreneurs. It will, however, increase investment for entrepreneurs conditional on entry. 

Overall, the steady state moments in the lottery economy are very similar to those in the benchmark economy. It could be that the model uses a discrete ability ($\theta$) grid so that not a lot of entrepreneurs are on the margin, or that the size of the lottery sector is not large enough.

% how to reconcile with literature
% discretized grid for asset and ability
\subsection{Connecting Model with Literature}
The model environment is closest to the empirical setting in \cite{bermejo_entrepreneurship_2018}. The paper documents the effect of winning lottery prizes from the Spanish Christmas Lottery on entrepreneurship. The Spanish Christmas Lottery is a national lottery game that takes place every year. It covers about 75\% of the population. Players are ordinary citizens, and the amount of money spend is similar across individuals. The only difference is that prizes are distributed to the entire province, and the winning province gets 5.65\% of its gross domestic product (GDP), a higher proportion of GDP than what the theoretical model assumes. The paper shows that lottery prizes have a positive effect on the rate of firm creation and wages. Moreover, conditional on entry, assets of firms are higher in winning provinces. Thus, the empirical findings are in line with the predictions of my model.

\cite{cespedes_more_2021} discusses a US setting, although the targeted population is the retail businesses. Retailers that sell jackpot tickets for Powerball and Mega Millions receive bonuses, which is a percentage of the winning prize, when they have sold winning tickets. Similar to model predictions, cash windfalls increase investments into existing business and spur new business creation. The advantage of distributing random bonuses to the retailers is that bonuses are more likely to be used for entrepreneurial activities, since retailers are already self-selected entrepreneurs with high enough entrepreneurial ability. That is, would-be entrepreneurs on the margin are more likely to be financially constrained rather than having low abilities. In the model, some prizes are distributed to workers with low entrepreneurial ability, and hence have no impact on entrepreneurial activities. The disadvantage of focusing on retailers is that the type of business started is likely limited to a certain industry. Thus, policymakers should weigh the pros and cons when considering the target population for similar cash transfer programs.

% implementation assumption and impact
% comparison with small business policy
\subsection{Policy Implications}
Results from the lottery economy have the following lessons for policy aimed at spurring entrepreneurship. If the US government would like to initiate a cash transfer program similar to the nationwide lottery program in Spain, the size of lottery prizes and total amount of cash being redistributed have to be large enough to overcome the fixed cost of starting a business. Secondly, the target population of the cash transfer program should cover those with entrepreneurial experience, so that they have high enough ability to start businesses after winning prizes. The cash transfer program will not work if the barrier to entry is due to a lack of entrepreneurial ability.

\section{Conclusion}
This chapter augments a standard Bewley model with an entrepreneurial sector and lottery sector to study how nationwide lottery programs like those implemented in Norway or Spain affect entrepreneurial dynamics and aggregate variables. The quantitative analysis shows that after winning a lottery prize, the likelihood of becoming an entrepreneur depends on the ability of would-be entrepreneurs on the margin. On the other hand, investments by existing entrepreneurs always increase after winning lottery prizes. Results of the model explain why public cash transfer programs may not increase the number of new businesses and only modestly increase aggregate productivity and output.

\section*{Appendix}
\subsection{Computation Algorithm}
\subsubsection{Benchmark Economy Steady State}
I use bisection on capital-labor ratio to clear the capital and labor markets. Given the non-linearity of the problem, I solve for the stationary equilibrium by value function iterations in a discretized state space. The asset space has 1000 discrete points and ranges from 0.01 to 20. Entrepreneurs choose capital and labor from 1000 points respectively.

\begin{itemize}
	\item Step 1: Guess a set of value functions for workers and entrepreneurs, capital-labor ratio in the corporate sector and use it to compute $r$ and $w$.
	\item Step 2: Given $r$ and $w$, solve the household optimization problem using the value function iteration and derive policy functions for each state.
	\item Step 3: Given the transition rules of asset and occupation derived in Step 2, compute the invariant distribution by discretizing the density function (histogram method).
	\item Step 4: Compute aggregate capital and labor using the invariant distribution derived in Step 3 and compute a new capital-labor ratio. Check if it is close to the initial guess of capital-labor ratio, if not, update the guess and go back to Step 2.
\end{itemize}

\subsubsection{Lottery Economy Steady State}
Use bisection method with the outer loop on targeted capital output ratio, and the inner loop on equilibrium interest rate. I use 100 points on the asset grid to speed up convergence, thus the wealth distribution is not that smooth. 
\begin{itemize}
	\item Step 1: Guess a set of value functions for workers and entrepreneurs, lottery ticket price, capital-labor ratio in the corporate sector and use it to compute $r$ and $w$.
	\item Step 2: Given $r$ and $w$, solve the household optimization problem using the value function iteration and derive policy functions for each state.
	\item Step 3: Given the transition rules of asset and occupation derived in Step 2, compute the invariant distribution by discretizing the density function (histogram method).
	\item Step 4: Compute aggregate capital and labor using the invariant distribution derived in Step 3 and compute a new capital-labor ratio. Check if it is close to the initial guess of capital-labor ratio, if not, update the guess and go back to Step 2.
	\item Step 5: Compute aggregate output using the invariant distribution derived in Step 3 and get a new lottery ticket price. Check if it is close to the initial guess of lottery ticket price, if not, update the guess and go back to Step 2.
\end{itemize}

\subsubsection{Simulation}
Initialize a panel of 400,000 households, 
\begin{itemize}
	\item Step 1: In period 1, initialize households' first period ability and efficiency, $\theta,\eta$, by randomly sampling from the stationary distribution of $\theta$ and $\eta$ respectively. Initialize every household to be a worker (so there will not be entrepreneurs with ability zero) with an asset endowment, $a$, at the steady state capital.
	\item Step 2: Starting in period 2, update households' $\theta$ and $\eta$ by drawing on the transition matrix, given last period's values. Distribute lottery prizes $\psi$ randomly by the winning probability distribution. Then update households' savings, consumption, occupation, and investment (if entrepreneur) choice, given $a,\theta,\eta,\psi$, using the policy functions solved in steady state.
	\item Step 3: repeat Step 2 and move forward until period 200. Check if aggregate capital converges.
\end{itemize}
I use the results from the last two periods for the regression.

%\textit{Computation of transition path}\\
%Assume the economy is in initial steady state in period 0 and lottery outcomes are announced from periods $t_1$ to $t_2$. The economy makes a transition to the final steady state in period $T$. Choose $T$ large enough so that the transition path is not affected by increasing $T$.
%\begin{itemize}
%	\item Step 1: Guess the interest rate path and update labor-capital ratios and wage path.
%	\item Step 2: Given prices, use the value function of the final steady state for the period $T$ and solve the households' problem backwards from period $T-1$.
%	\item Step 3: Use the distribution of the initial steady state and the policy functions from step 2 to compute forward the path of the distribution.
%	\item Step 4: Compute the path of aggregate labor and capital using the distribution in step 3. Calculate excess labor demand, adjust the interest rate path and go back to step 2.
%\end{itemize}

\subsection{Markov processes for $\eta$ and $\theta$}
The Markov process for the labor productivity $\eta$ and $\theta$ are as follows: 

$\eta$ grid = $[0.646, 0.798, 0.966, 1.169, 1.444]$, \\

transition matrix $P_\eta$=
$\begin{bmatrix}
	0.731 & 0.253 & 0.016 & 0.000 & 0.000 \\
	0.192 & 0.555 & 0.236 & 0.017 & 0.000 \\
	0.011 & 0.222 & 0.533  & 0.222 & 0.011 \\
	0.000 & 0.017 & 0.236 &  0.555 & 0.192 \\
	0.000& 0.000 & 0.016 & 0.253 & 0.731
\end{bmatrix}$\\

stationary distribution = $[0.166, 0.218, 0.232, 0.218, 0.166]$.\\

$\theta$ grid = $
[0.000, 0.706, 1.470, 2.234]$,\\

transition matrix $P_\theta$=
$\begin{bmatrix}
	0.780& 0.220 & 0.000 & 0.000 \\
	0.430& 0.420& 0.150 & 0.000 \\
	0.000& 0.430& 0.420& 0.150 \\
	0.000& 0.000& 0.220& 0.780 \\
\end{bmatrix}$\\

stationary distribution = $[0.554, 0.283, 0.099, 0.064]$.

\subsection{Gambling in Norway}
Only the two state-owned entities, Norsk Tipping and Norsk Rikstoto, are allowed to offer gambling services. The aim of these two companies is to provide entertainment within responsible limits, with the profits going to charitable causes. Unregulated gaming accounts for 14.0\% of the total gambling market. According to Norsk Tipping's 2012 Annual Report, 70\% of Norwegians above the age of 18 gambled in 2012 through their services. In 2016, customer growth has increased though both physical and digital distribution channels. Lottery gaming accounts for the largest share of sales, and represents 65\% of the gross gaming revenues in 2016.

It is important to understand why people buy lotteries to identify any potential difference from the general population. Norsk Tipping claims that its task is to channel Norwegians' desire to play to legal, regulated games in Norway. It advertises its games as fulfilling the majority of customers' entertainment needs while counteracting the negative impacts of irresponsible gambling behavior. In 2014, Norsk Tipping introduced applications that provide players with improved insight into their own gaming habits and conduct. Thus, it is reasonable to view gambling through these two companies as a pure entertainment activity. The risk-loving is more likely to self-select into the unregulated market.

% *****************************************************************
% BIBLIOGRAPHY
%******************************************************************
\begin{singlespace}

	\bibliographystyle{econometrica}
	\bibliography{thesis_vc,masterbib}
\end{singlespace}

\end{document}